%% file: main.tex
\documentclass[10pt]{iopart}

\usepackage{siunitx}%[output-decimal-marker={,}]
\usepackage{amsfonts}
\usepackage{amssymb}
\usepackage{graphicx}
\usepackage[autostyle]{csquotes} % required for \enquote{}
\usepackage{bm}
\usepackage{subcaption} % provides subfigure environment
\usepackage{booktabs}
\usepackage{makecell}
\usepackage[switch]{lineno} % for line numbering using the \linenumbers command
\DeclareCaptionFormat{custom}
{%
    \textbf{#1#2}\textit{#3}
}

\usepackage[
    colorlinks=true,
    linkcolor=blue,
    citecolor=blue,
    urlcolor=blue,
    pdfauthor={Sascha Rienaecker},
    bookmarksopen=true,
    ]{hyperref}
    
\usepackage[backend=biber,
    bibstyle=numeric-comp,
    citestyle=numeric-comp, %
    maxcitenames=1,
    sorting=none, % sorting=ynt (year, name title)
    giveninits=true,
    ]{biblatex}

\usepackage{biblatex-shortfields}
\usepackage{datetime}
\newdateformat{monthyear}{\monthname[\THEMONTH] \THEYEAR}

\AtEveryBibitem{%
\clearfield{url}%
\clearfield{urlyear}%
\clearfield{title}%
}

\newcommand{\low}[1]{_\mathrm{#1}}
\newcommand{\up}[1]{\mathrm{#1}}

\newcommand{\refig}[1]{Fig.\,\ref{fig:#1}}
\newcommand{\refsec}[1]{Sec.\,\ref{sec:#1}}
\newcommand{\reftab}[1]{Tab.\,\ref{tab:#1}}

\def\documentspath{/home/rienacker/Documents} % temporary path, figures should be placed later in the root directory of the main.tex file instead

\def\EPSfigpath{\documentspath/EPS24-Salamanca/figures}

\newcommand{\ErxB}{E_r \times B}
\newcommand{\BxgradB}{B \! \times \! \nabla B}

\graphicspath{
              {\EPSfigpath/}
              }

\usepackage[final, authormarkup=none=color, commandnameprefix=ifneeded]{changes} % Use option final to hide changes
\makeatletter
\newcommand{\mainmatter}{%
\setcounter{footnote}{0}%
\patchcmd{\@makefntext}{\fnsymbol}{\alph}{}{}%
\patchcmd{\@thefnmark}{\fnsymbol}{\alph}{}{}%
\def\@makefnmark{\textsuperscript{\alph{footnote}}}%
}
\makeatother

\begin{document}

\title{Edge Radial Electric Field in Positive and Negative Triangularity Plasmas in the TCV Tokamak}

%%% So, we use this manual list instead:
\author{S. Rienäcker$^1$, P. Hennequin$^1$, L. Vermare$^1$, C. Honoré$^1$, R. Bouffet-Klein$^2$,
        S. Coda$^3$, B. Labit$^3$, B. Vincent$^3$, K.E. Thome$^4$, O. Krutkin$^3$, A. Balestri$^3$, Y. Nakeva$^{5}$ and 
        the TCV team\footnote{See author list of B.P. Duval et al 2024 Nucl. Fusion 64 112023}}

\address{$^1$ Laboratoire de Physique des Plasmas (LPP), CNRS, Sorbonne Université, École polytechnique, Institut Polytechnique de Paris, Palaiseau, France}
\address{$^2$ ENS Paris-Saclay, Gif-sur-Yvette, France}
\address{$^3$ Ecole Polytechnique Fédérale de Lausanne, Swiss Plasma Center, Lausanne, Switzerland}
\address{$^4$ General Atomics, San Diego, CA, United States of America}
\address{$^5$ Università degli Studi della Tuscia, Viterbo, Italy}

\ead{sascha.rienacker@lpp.polytechnique.fr}
\vspace{10pt}
\begin{indented}
\item[]\monthyear\today 
\end{indented}

% \linenumbers % Start line numbering

% \begin{abstract}
% \ldots
\input{abstract.tex}
% \end{abstract}

% \submitto{\NF}
%\maketitle
\ioptwocol

\mainmatter

\input{sections/introduction.tex}

\input{sections/method.tex}

\input{sections/results.tex}

\input{sections/conclusion.tex}

\appendix

\input{sections/ack.tex}

\printbibliography

\end{document}

%% file: abstract.tex
\begin{abstract}
    We present the first edge $E_r$ measurements in negative triangularity (NT) TCV plasmas. 
    The Doppler backscattering measurements of $v_\perp \approx E_r/B$ reveal a significant impact of triangularity on the $E_r$ well:
    In Ohmic, NBI, and ECRH heated discharges, the $E_r$ well and associated $\ErxB$ shear are stronger in NT-shaped plasmas compared to their positive triangularity (PT) counterpart.
    This suggests a connection to the concomitant NT performance gain relative to PT L-mode.
\end{abstract}

%% file: sections/introduction.tex
\section{Introduction}\label{sec:intro}

Efforts to reconcile high fusion performance in future tokamak reactors with power handling demands have motivated research into non-conventional plasma confinement modes and configurations\,\cite{Viezzer_2023}.
Negative triangularity (NT) has emerged in recent years as a promising alternative to the conventional positive triangularity (PT) plasma geometry (see \cite{Marinoni_2021} for a review), aligned with a \enquote{power-handling-first} reactor design strategy\,\cite{Kikuchi_2019}.
By preventing the transition from L- to H-mode, NT passively suppresses edge-localized modes (ELMs)---harmful instabilities associated with H-mode that future reactors must avoid\,\cite{Viezzer_2023}. At the same time, NT-shaped plasmas exhibit particularly good L-mode confinement\,\cite{Camenen_2007, Austin_2019, Marinoni_2019,Coda_2021, Happel_2022, Thome_2024, PazSoldan_2024, Balestri_2024}, comparable with that of PT H-mode.
These beneficial features of NT provide strong motivation to understand their underlying mechanisms.
A prominent hypothesis for the H-mode suppression in sufficiently NT shapes invokes the destabilization of infinite-$n$ MHD ballooning modes, closing access to second stability\,\cite{Merle_2017,Saarelma_2021, Nelson_2022}.
Concerning the outstanding L-mode confinement observed in NT, while a generally conclusive picture has not emerged yet, several relevant hints have been obtained: The empiric reduction of fluctuation levels in NT compared to PT\,\cite{Fontana_2017,Huang_2018,Austin_2019,Marinoni_2019, Fontana_2019, Han_2021, Krutkin_2024, Stewart_2025}, as well as modeling results\,\cite{Marinoni_2009, Merlo_2021, Merlo_2023, DiGiannatale_2024, Balestri_2024b, Garbet_2024, Ulbl_2025_preprint}, point to a beneficial effect of NT on ion-scale turbulence. In particular, modeling results suggest a stabilization of trapped electron modes (TEM)\,\cite{Marinoni_2009, Garbet_2024, Ulbl_2025_preprint}, while recent studies find a similar impact also on ion temperature gradient (ITG) driven modes\,\cite{Balestri_2024b, Merlo_2023}.
Further, experimental findings from DIII-D\,\cite{Nelson_2024} and TCV\,\cite{Sauter_2014, Balestri_2024} highlight the importance of the plasma edge, attributing the performance gain in NT partly to increased values of pressure and of its gradient near the separatrix. Yet, the fundamental reason for this distinct edge behavior remains unclear.
Crucial to edge transport is the radial electric field ($E_r$) \enquote{well} that forms within a thin radial layer just inside the separatrix. The associated $\ErxB$ velocity shear is widely recognized as playing a major role in the regulation of turbulent transport\,\cite{Biglari_1990, Burrell_1997} and, consequently, of confinement.
While the tokamak edge and in particular its $E_r$ structure remain generally insufficiently understood, this is especially true for NT.
Characterizing the NT edge and comparing it to PT with a particular focus on $E_r$ might offer insights into tokamak edge transport in general, which is key to improving predictive confinement models for future devices.

In this context, we report on Doppler backscattering measurements of the edge $E_r$ profile in NT compared to PT shaped plasmas in the \textit{Tokamak à Configuration Variable} (TCV)\,\cite{Duval_2024} (major radius $R = \SI{0.88}{m}$, 
    minor radius $a = \SI{0.25}{m}$, on-axis magnetic field $|B_0| < \SI{1.54}{T}$, plasma current $|I_p| < \SI{1}{MA}$).
This study extends the scarce database on the $E_r$ well in NT\,\cite{Happel_2022, Vanovac_2024, Stewart_2025}. Our data provide what is, to our knowledge, the first systematic comparison of $E_r$ in matched NT/PT discharges, isolating the effect of triangularity.
The structure of this letter is as follows: \refsec{method} outlines the diagnostics and analysis approach. \refsec{resultsA} discusses NT/PT comparisons in matched L-mode discharges, and \refsec{resultsB} addresses higher-performance scenarios. A summary is provided in \refsec{conclusion}.

%% file: sections/method.tex
\section{Experimental Method}\label{sec:method}

As a proxy for $E_r$ (the main focus of this study), we use the turbulence perpendicular velocity $v_\perp$ measured via Doppler backscattering (DBS)\,\cite{Hirsch_2001, Hennequin_2006}.
DBS uses a microwave beam to probe the plasma and detects the wave scattered back by density fluctuations. The scattering fluctuations are localized near the beam turning point and have a perpendicular wave number, $k_\perp$, selected by the scattering geometry.
The Doppler shift ($\omega\low{D})$ introduced by the moving scatterers gives access to their lab frame velocity, $v_\perp = \omega\low{D} / k_\perp$, which corresponds to the $\ErxB$ rotation if the turbulence intrinsic velocity is negligible. $v_\perp$ is inferred by extracting $\omega\low{D}$ from the power spectral density (PSD) of the DBS signal, combined with beamtracing to determine the beam turning point and the selected $k_\perp$.
Radial profiles of $v_\perp$ (or $E_r$) are obtained by stepping the probing frequency.

TCV is currently equipped with a dual-channel V-band DBS diagnostic, on loan from LPP\,\cite{Rienaecker_2025}. It allows edge $v_\perp$ profiles to be measured within a typical acquisition time of \SI{100}{ms}.
The DBS beam is launched from the upper low field side, providing access to the first quadrant of the plasma cross-section, see \refig{85429_85430_eq}. 
The DBS system and data processing method at TCV are described in Ref.\cite{Rienaecker_2025}.
All radial profiles in this letter are displayed as a function of the flux surface label, $\rho_\psi$, defined as the square root of the normalized poloidal flux. The sign of $v_\perp$ is chosen positive (negative) when pointing in the ion (electron) diamagnetic direction. Otherwise, we adopt the standard COCOS17\,\cite{Sauter_2013} toroidal coordinates convention (a positive $I_p$ or $B_0$ is anti-clockwise when viewed from above).
    
Besides $v_\perp$, the analysis includes electron density ($n_e$) and temperature ($T_e$) profiles constructed from Thomson scattering (TS)\,\cite{Blanchard_2019} data. 
Information on the ion temperature ($T_i$) is obtained by charge exchange recombination spectroscopy (CXRS)\,\cite{Bagnato_2023} using the CX reactions triggered by a diagnostic neutral beam (DNBI)\,\cite{Karpushov_2009} on carbon impurity (C$^{6+}$) ions. 
The CXRS profiles are also used for an estimation of the outer core profile of $E_r$ through the radial force balance, following the procedure described in Ref.\,\cite{Rienaecker_2025}: The poloidal impurity flow is computed using the neoclassical code NEO\,\cite{Belli_2008} with kinetic profiles as inputs, while the diamagnetic and toroidal velocity contributions to the force balance are obtained from spline fits to the CXRS data. 
The diagnostics' probing locations are indicated in \refig{85429_85430_eq}.

Confinement level is quantified using conventional 0D metrics:
the steady-state ($\mathrm{d}W/\mathrm{d}t=0$) energy confinement time, $$\tau_E = W/P\low{in}\, ,$$ 
the H-mode enhancement factor (H-factor), $$H\low{98y2} = \tau_E / \tau_E^{\mathrm{IPB98(y,2)}}$$
and the normalized beta, $$\beta_N = \beta \up{[\%]} a \up{[m]} B_0 \up{[T]} / I_p \up{[MA]}\, .$$
In these expressions, $W$ is the total thermal energy content computed from the (electron + ion) pressure profile $p=p_e+p_i$ (inferred from TS and CXRS data).
$P\low{in}$ is the injected power, $\tau_E^{\mathrm{IPB98(y,2)}}$ the IPB98(y,2) scaling law\,\cite{ITER_1999} and $\beta=2 \langle p \rangle \mu_0/ B_0^2$ the thermal to magnetic pressure ratio.

Finally, we characterize the magnetic geometry using the edge safety factor $q_{95}$, elongation $\kappa$, bottom and top triangularities $\delta\low{bot/top}$, and their average $\overline{\delta}$, as well as the ion $\BxgradB$ drift direction (termed \enquote{favorable} or \enquote{unfavorable} when pointing towards or away from the X-point, respectively).

%% file: sections/results.tex
\section{NT vs. PT in matched L-mode conditions}\label{sec:resultsA}

\begin{figure*}[hbtp]
    \centering
    \begin{minipage}[b]{0.19\linewidth}
        \centering
            {\footnotesize
            \setlength{\tabcolsep}{3pt} % Reduce horizontal padding
            \begin{tabular}{lc}
                \toprule
                & \textcolor{red}{PT} / \textcolor{blue}{NT} \\
                \midrule
                $P\low{Ohm}$\,[kW]     & \multicolumn{1}{c}{195} \\
                $I_p$\,[kA]     & \multicolumn{1}{c}{170} \\
                $B_0$\,[T]      & \multicolumn{1}{c}{1.44} \\
                $\kappa$        & \multicolumn{1}{c}{1.5}  \\
                $\delta\low{top}$        & \textcolor{red}{0.22}/\textcolor{blue}{0.25}   \\
                $\delta\low{bot}$        & \textcolor{red}{0.32}/\textcolor{blue}{-0.34}   \\
                $q_{95}$                 & \textcolor{red}{4.2} /\textcolor{blue}{3.9}    \\
                $\tau_E$ [ms]            & \textcolor{red}{22} /\textcolor{blue}{28}    \\
                $H\low{98y2}$            & \textcolor{red}{0.87} /\textcolor{blue}{1.04}    \\
                $\beta_N$                & \textcolor{red}{0.83} /\textcolor{blue}{1.0}    \\
                \bottomrule 
                % \vspace{2cm}
            \end{tabular}
            }
            \captionof{table}{Discharge parameters of the matched NT/PT pair.}\label{tab:USN_NT_PT}
        \end{minipage}
    \begin{minipage}[t]{0.25\linewidth}
        \centering
        \includegraphics[width=\linewidth]{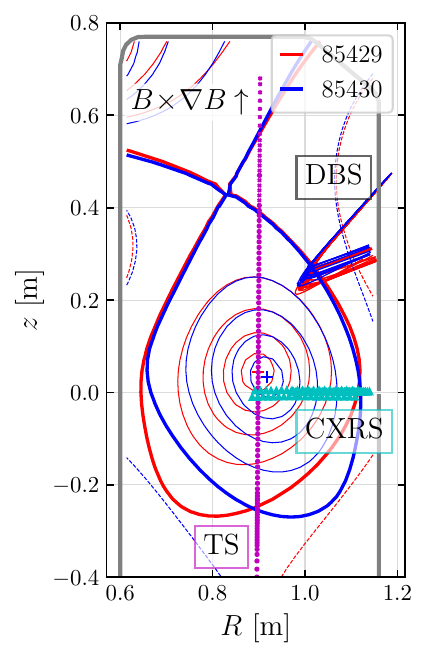}
        \caption{Contours of $\rho_\psi$ for the two shapes, with DBS beam paths and TS/CXRS diagnostic locations indicated.}\label{fig:85429_85430_eq}
    \end{minipage}
    \hspace{0.25cm}
    \begin{minipage}[t]{0.52\linewidth}
        \centering
            \includegraphics[width=\linewidth]{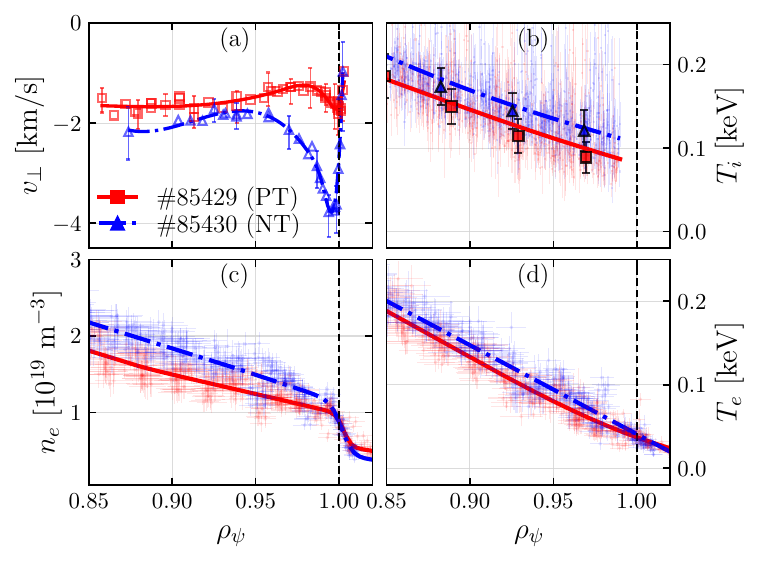}
            \caption{Edge profiles acquired over a stationary phase $t\in[1.0,1.5]$s: (A) Turbulence ($\ErxB$) velocity from DBS, (B) impurity (C${^6+}$) ion temperature from CXRS, (C) density and (D) electron temperature from TS. Fits are overlaid to guide the eye; for $T_i$, markers also show the binned weighted average and standard deviation.}\label{fig:85429_85430_profiles}
        \end{minipage}
        % \caption{}\label{fig:85429_85430}
\end{figure*}

% discharge conditions and equilibria
First, a pair of upper single-null (USN), Ohmic L-mode deuterium discharges with favorable $\BxgradB$ drift is investigated. The shapes have opposite edge \textit{bottom} triangularity ($\delta\low{bot}\approx\pm 0.3$), but the equilibria are largely matched otherwise (\refig{85429_85430_eq}). The main discharge parameters are summarized in \reftab{USN_NT_PT}. Note that at fixed $I_p$ and $B_0$, the NT geometry results in lower $q_{95}$ than PT. The possible effect of $q_{95}$ on $E_r$\,\cite{Vermare_2021,Varennes_2023} is not discussed here.
Keeping the upper half of the shapes fixed ensures nearly identical DBS probing geometries. This minimizes possible biases e.g. due to differences in local flux expansion or probed turbulence wave numbers ($k_\perp$). The latter varies radially in the range of $k_\perp \in [5,6] \, \SI{}{rad \per \centi \m}$ ($k_\perp \rho_s \sim 0.5$)\footnote{The ion Larmor radius at soundspeed, $\rho_s$, is evaluated using $T_e (\rho_\psi=0.95) \approx \SI{0.1}{keV}$.}, but differs only slightly ($\lesssim 10\%$) between NT and PT.

The $v_\perp$ profiles corresponding to a stationary time window are displayed in \refig{85429_85430_profiles}\,(A), $T_i$ in (B), and $n_e$ and $T_e$ in (C) and (D), respectively.

A remarkable difference is observed: The NT case exhibits a sharp well of $v_\perp^{\up{min}} \approx -\SI{4}{km/s}$, while that of the PT case has a depth $v_\perp^{\up{min}} \approx \SI{-2}{km/s}$, typical of comparable L-mode PT discharges in TCV\,\cite{Rienaecker_2025}.
The kinetic profiles reveal a higher L-mode density pedestal in NT, radially coinciding with the velocity well, where $v_\perp$ differs most between NT and PT. 
At same levels of Ohmic heating, $\tau_E$ is increased by roughly $\SI{30}{\%}$ in the NT case (see \reftab{USN_NT_PT}), as reflected by higher density---despite lower fueling---and slightly higher ion and electron temperatures, up to the core (not shown). 
Meanwhile, the edge C$^{6+}$ toroidal velocity ($v_\varphi$, not shown) does not differ notably between NT and PT within experimental uncertainties. In the core, an intrinsic counter-$I_p$ rotation of $v_\varphi \approx 15 \pm \SI{5}{km/s}$ is observed (slightly weaker in PT), decaying toward near-zero values at the edge.

\begin{figure}[htbp]
    \centering
    \includegraphics[width=1\linewidth]{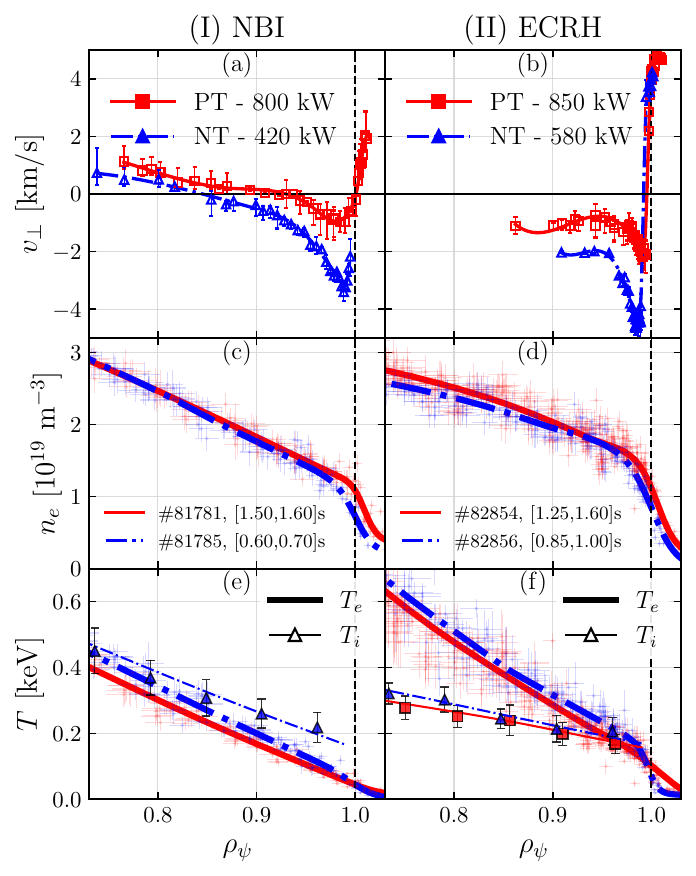}
    \caption{(a-b) Turbulence $\ErxB$ velocity and (c-f) edge kinetic profiles for equilibria close to those in \refig{85429_85430_eq} but with auxiliary heating: NBI (left column) and ECRH (right column), using higher power in PT for a better match of the electron kinetic profiles. Due to missing CXRS data in this interval, $T_i$ profiles are omitted for \#81781.}\label{fig:NT_PT_USN_NBI_and_ECRH_matched_profiles}
\end{figure}

\begin{figure*}[htbp]
    \centering
    \includegraphics[width=0.95\linewidth]{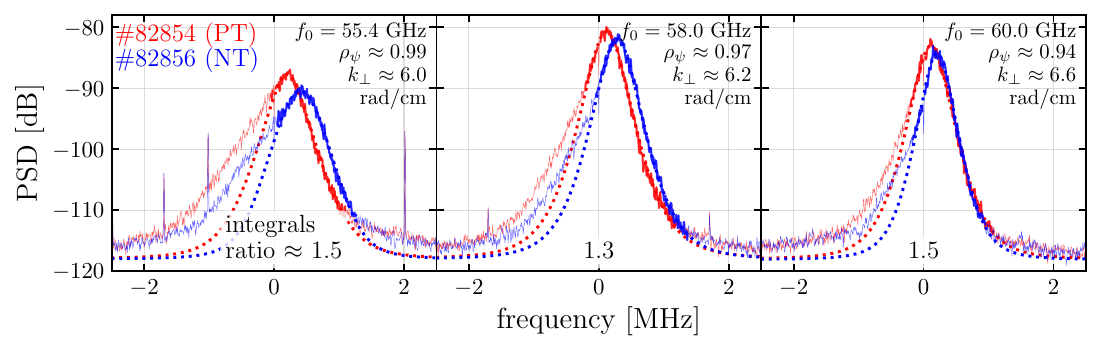}
    \caption{DBS power spectral density (PSD) comparison between NT and PT at matched density (ECRH scenarios from \refig{NT_PT_USN_NBI_and_ECRH_matched_profiles}). Dashed lines indicate fits to the main peak, whose integral reflects the turbulence intensity $|\tilde{n}(k_\perp)|^2$. The PT to NT ratio of this integral is annotated at the bottom. At fixed probing frequencies ($f_0$) between NT and PT, $\rho_\psi$ and $k_\perp$ are matched to within \SI{1}{\%} and \SI{2}{\%}, respectively, allowing for a meaningful comparison of $|\tilde{n}(k_\perp)|^2$.}\label{fig:NT_PT_USN_ch2_spectra}
\end{figure*}

The trend of improved confinement and a deeper $v_\perp$ well in NT persists under moderate auxiliary heating, whether by co-current neutral beam injection (NBI) or second-harmonic electron cyclotron resonance heating (ECRH). This behavior is observed both at equal heating powers---$P_\text{NBI} = \SI{420}{kW}$ and $P_\text{ECRH} = \SI{580}{kW}$ (not shown)---and when the heating power is adjusted between NT and PT to achieve matched kinetic profiles, as shown in \refig{NT_PT_USN_NBI_and_ECRH_matched_profiles} and further discussed below.
This suggests a degree of robustness, particularly with respect to external momentum input and to the dominant heating species (leading to $T_i \sim T_e$ or $T_i \ll T_e$ in the core for NBI and ECRH, respectively).
\newline
For an analysis of the turbulence level using DBS data, we examine two NT/PT pairs with profiles shown in \refig{NT_PT_USN_NBI_and_ECRH_matched_profiles}. They are heated by NBI and ECRH, respectively, with higher power applied in PT to better match the NT kinetic profiles.
The matched density profiles (\refig{NT_PT_USN_NBI_and_ECRH_matched_profiles}c--d) enable a qualitative assessment of the turbulence intensity, around a given $k_\perp$\,\cite{Hennequin_2006}. More precisely, at fixed probing frequencies, the DBS signals' power provide a relative measure of $|\tilde{n}(k_\perp)|^2$ between NT and PT, at a common $\rho_\psi$ and $k_\perp$.
In the two pairs considered in \refig{NT_PT_USN_NBI_and_ECRH_matched_profiles}, the DBS power tends to be higher in PT by a factor varying between $\sim 1$--$1.8$.
This is exemplified for the ECRH pair in \refig{NT_PT_USN_ch2_spectra}, where the PSD is compared between NT and PT for a few representative probing frequencies.
The suggested reduction of ion-scale edge turbulence in NT appears consistent with previous experimental findings for matched NT/PT profiles\,\cite{Huang_2017_phdthesis,Fontana_2019} and the modeling results\,\cite{Marinoni_2009, Merlo_2021, Merlo_2023, DiGiannatale_2024, Balestri_2024b, Garbet_2024, Ulbl_2025_preprint} mentioned earlier. % see section 3.3.2 of Huang's PhD Thesis
Further analyses are foreseen to assess how systematic the trend is and how it correlates to differences in the $E_r$ profile.

\section{Towards higher performance}\label{sec:resultsB}

\begin{figure*}[htbp]
    \centering
    \begin{minipage}[c]{0.49\textwidth} % Width for the figure
        \centering
        \includegraphics[width=\linewidth]{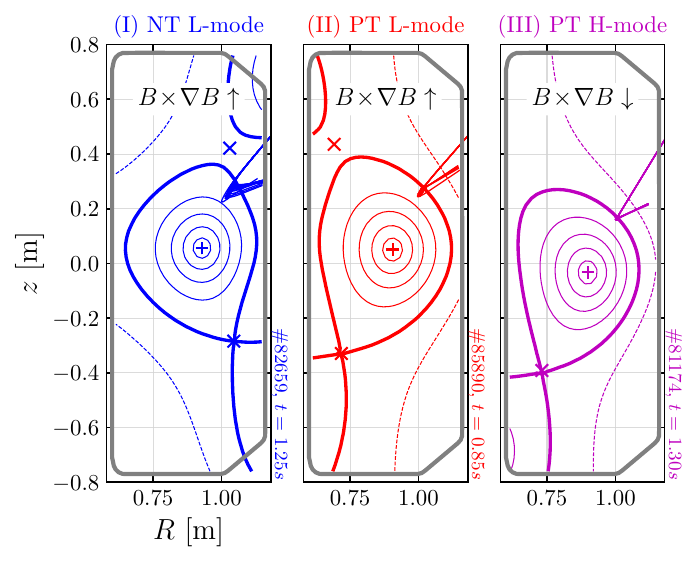}
        \caption{Plasma shapes of the higher performance scenarios.}\label{fig:NT_Lmode_vs_PT_Hmode_eq}
    \end{minipage}%
    \hfill
    \begin{minipage}[c]{0.49\textwidth} % Width for the caption
        \centering
        % \caption{\textsuperscript{\ref{cxrs_note}}}\label{tab:NT_L_PT_H}
        \begin{tabular}{l| >{\color{blue}}c| >{\color{red}}c| >{\color{magenta}}c}
            \toprule
            & NT L & PT L & PT H \\
            \midrule
            $\#$                    &    82659  &  85890    &   81174   \\      
            $t$ window [s]          &    [1.2,1.3]  &  [0.7,0.9]    &   [1.2,1.4]   \\      
            $P\low{NBI}$\,[kW]      &    \textcolor{black}{850}    &      \textcolor{black}{850}  &   \textcolor{black}{850}     \\
            $I_p$\,[kA]             &     $+\textcolor{black}{150}$    &   $+\textcolor{black}{150}$      &  $-\textcolor{black}{150}$         \\ 
            $B_0$\,[T]              &     $+\textcolor{black}{1.44}$       &  $+\textcolor{black}{1.44}$         & $-\textcolor{black}{1.44}$          \\ 
            $\kappa$                &      1.35 &   1.48    &    1.51 \\ 
            $\overline{\delta}$     &     $-0.49$&   0.52   &   0.48  \\      
            $q_{95}$                &       4.0 &     5.7   &    5.1  \\
            $\tau_E$ [ms]           &       8.6 &      6.7   &    14  \\
            $H\low{98y2}$           &      0.98 &    0.65   &    1.2  \\
            $\beta_N$               &      1.6  &    1.1    &    2.2  \\
            % $\fG$                   &      0.34 &    0.39   &    0.41  \\
            \bottomrule
        \end{tabular}
        \captionof{table}{Overview of main discharge parameters.\textsuperscript{\ref{cxrs_note}}}\label{tab:NT_L_PT_H}
        \end{minipage}
\end{figure*}

Next, we investigate how the $E_r$ comparison in NT vs. PT extends towards higher performance, which is more representative of conditions in future reactors. The NT equilibrium is shown in \refig{NT_Lmode_vs_PT_Hmode_eq}-I. Notably, it has a fully developed negative triangularity, $\overline{\delta} \approx -0.5$, and unfavorable $\BxgradB$ drift.
Subsequent results suggest that even with an unfavorable $\BxgradB$ direction---a configuration usually associated with a shallow edge $E_r$ in L-mode\,\cite{Schirmer_2006,Vermare_2021,Plank_2023}---NT maintains a sharp $E_r$ well along with good L-mode performance.

\begin{figure*}[htbp]
    \begin{subfigure}[t]{\linewidth}
        \raggedleft
        \includegraphics[width=\linewidth]{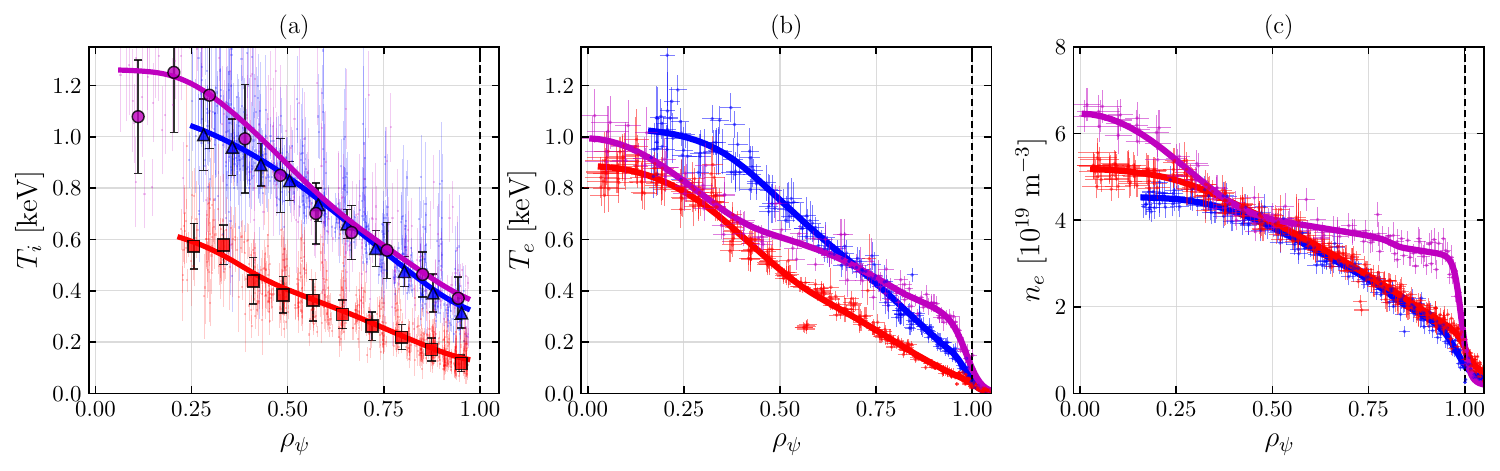}
    \end{subfigure}
    \caption{Kinetic profiles for the cases shown in \refig{NT_Lmode_vs_PT_Hmode_eq}, using inter-ELM data in H-mode: (a) impurity ion temperature from CXRS, (b) electron temperature and (c) density from TS.}\label{fig:NT_Lmode_vs_PT_Hmode_kinetic_profiles}
\end{figure*}

At fixed auxiliary heating power ($P\low{NBI} = \SI{850}{kW}$) delivered by a co-current NBI beam\footnote{Note that since $I_p<0$ in the PT H-mode and $I_p>0$ in the NT and PT L-mode discharges, different neutral beam sources are used (NBI-1 and NBI-2, respectively).}, the NT scenario is confronted  both to its mirrored PT L-mode counterpart (\refig{NT_Lmode_vs_PT_Hmode_eq}-II), and to a comparable PT H-mode discharge (\refig{NT_Lmode_vs_PT_Hmode_eq}-III) with favorable $\BxgradB$ drift.
While the PT L-mode scenario is useful for isolating the effect of triangularity, the PT H-mode provides a representative, reactor-relevant benchmark for evaluating the NT edge behavior. The externally controlled parameters of the PT H-mode are largely comparable, although not exactly identical: Its geometry and the signs of both $I_p$ and $B_0$ differ from the NT/PT L-mode cases, but the magnitudes of $\overline{\delta}$, $I_p$ and $B_0$ are consistent.
The discharge parameters are summarized in \reftab{NT_L_PT_H}.
The corresponding kinetic profiles are displayed in \refig{NT_Lmode_vs_PT_Hmode_kinetic_profiles}.\footnotemark\label{cxrs_note}
\footnotetext{Owing to the lack of CXRS data in \#85890 (PT L-mode), the $T_i$ profile (\refig{NT_Lmode_vs_PT_Hmode_kinetic_profiles}) and confinement metrics $\tau_E$, $H\low{98y2}$ and $\beta_N$ (\reftab{NT_L_PT_H}) are taken from an equivalent discharge phase in \#84932 ($t\in[0.65,0.8]$\,s), which lacks usable DBS data.}


The PT H-mode exhibits quasi-regular type-I ELMs, while the other two discharges are entirely ELM-free, consistent with L-mode.
Moreover, the PT H-mode features a pronounced pedestal in $n_e$ and $T_e$ (\refig{NT_Lmode_vs_PT_Hmode_kinetic_profiles}b--c) which sets it apart from the L-mode cases. The NT discharge displays an L-mode-like edge density profile, but with respect to the PT L-mode, the NT edge temperature is notably higher, both in terms of $T_e$---in line with the trend observed in DIII-D\,\cite{Nelson_2024}, as well as $T_i$.
Judging by its kinetic profiles, and by the confinement metrics listed in \reftab{NT_L_PT_H}, the NT scenario shows an intermediate performance between the PT L- and H-mode.
Yet, due to markedly different edge conditions---notably in density---any comparison of $E_r$ or confinement level between L- and H-mode remains inherently limited. Furthermore, the higher overall density in the present H-mode tends to overstate its performance relative to the L-modes.

\begin{figure}[htbp]
    \centering
    \includegraphics[width=0.9\linewidth]{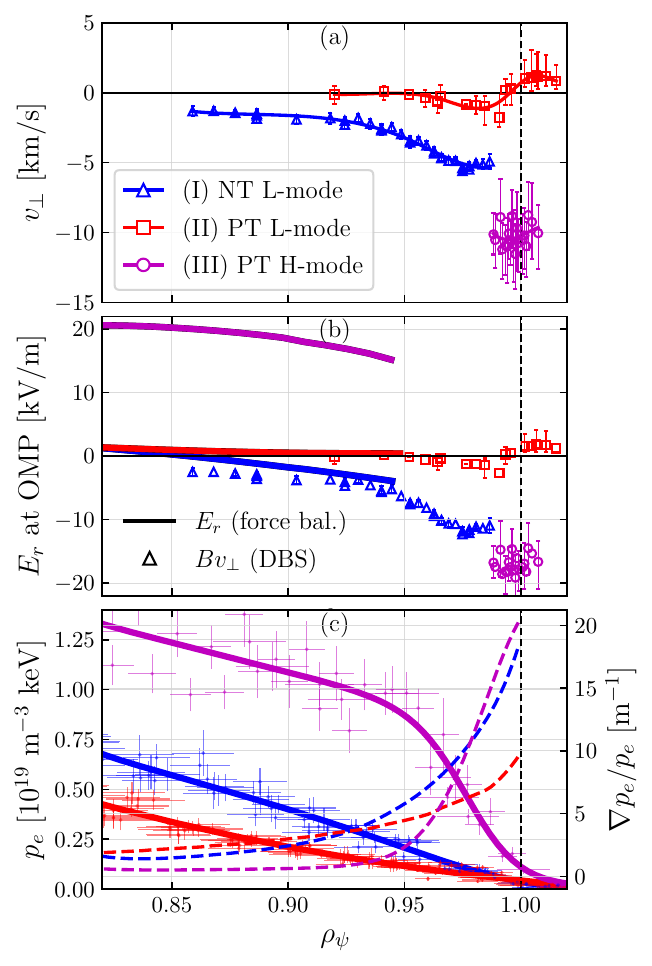}
    \caption{Edge profile comparison for the scenarios in \refig{NT_Lmode_vs_PT_Hmode_eq}: (a) local DBS velocity, (b) DBS data expressed in terms of $E_r$ and mapped to the OMP with the force balance result, (c) $p_e$ and its normalized gradient (solid/dashed).}\label{fig:NT_Lmode_vs_PT_Hmode_vperp}
\end{figure}

The edge $v_\perp$ profiles are shown in \refig{NT_Lmode_vs_PT_Hmode_vperp}\,(a).
As anticipated, the NT discharge exhibits a deeper well compared to its L-mode PT counterpart, consistent with the results of \refsec{resultsA}.
The associated $\ErxB$ shear around $\rho_\psi \approx 0.95$--$0.98$ is visibly stronger in NT.
In the PT H-mode, the low turbulence level and steep density pedestal limit the usable DBS data to a narrow radial range near the well minimum, preventing an assessment of the edge velocity shear from DBS alone.
We therefore resort to a complementary estimation of the outer core $E_r$ from the C$^{6+}$ radial force balance. Its profile is displayed in \refig{NT_Lmode_vs_PT_Hmode_vperp}b, along with the DBS data which is expressed in terms of $E_r = v_\perp B$ and mapped to the OMP. The consistency between the two $E_r$ estimations in the L-mode cases lends confidence to the analysis procedure.
The jump in $E_r$ in the PT H-mode between $\rho_\psi\approx 0.95$ and $1$ (\refig{NT_Lmode_vs_PT_Hmode_vperp}b) indicates a strong shear layer---as expected for H-mode---coinciding with the steep region of the $p_e$ pedestal (\refig{NT_Lmode_vs_PT_Hmode_vperp}c).
Overall, we conclude that, in the NT discharge, the $E_r$ well depth and associated velocity shear lie in between the PT L-mode and H-mode. Under favorable $\BxgradB$ drift, the L-mode $E_r$ well is expected to deepen further\,\cite{Schirmer_2006,Vermare_2021,Plank_2023}, possibly bringing the NT well depth closer to H-mode; this will be explored in future work.

%% file: sections/conclusion.tex
\section{Summary and conclusions}\label{sec:conclusion}

We have investigated the effect of triangularity on the edge $E_r$ well in TCV. 
In carefully matched NT/PT Ohmic discharges, NT unambiguously displays a deeper $E_r$ well accompanied by better confinement compared to its PT counterpart.
The trend persists upon moderate auxiliary heating using ECRH or NBI.
In a higher performance scenario, NT displays a remarkable L-mode $E_r$ well depth, intermediate between a comparable PT L-mode and H-mode. 
These observations confirm the distinct behavior of the NT edge\,\cite{Nelson_2024} now also in terms of the $\ErxB$ flow---a key element to a better understanding of edge transport and overall confinement in tokamaks.
The correlation observed between stronger $E_r \times B$ shear and confinement suggests a causal link consistent with the turbulence stabilization paradigm by sheared flows\,\cite{Biglari_1990, Burrell_1997}.
These observations motivate edge transport modelling to explore the link between shaping, edge flow shear and turbulence regulation, as well as deeper investigations into the edge turbulence properties of NT vs. PT.

%% file: sections/ack.tex
\ack

The main author would like to thank X. Garbet, P. Donnel, G. Durr-Legoupil-Nicoud, Y. Camenen and A. Merle for fruitful discussions and valuable input.
This work has been carried out within the framework of the EUROfusion Consortium, partially funded by the European Union via the Euratom Research and Training Programme (Grant Agreement No 101052200 — EUROfusion). 
This work has benefited from a grant managed by the Agence Nationale
de la Recherche (ANR), as part of the program \enquote{Investissements
d'Avenir} under the reference (ANR-18-EURE-0014). 
The Swiss contribution to this work has been funded by the Swiss State Secretariat for Education, Research and Innovation (SERI). Views and opinions expressed are however those of the author(s) only and do not  necessarily reflect those of the European Union, the European Commission or SERI. Neither the European Union nor the European Commission nor SERI can be held responsible for them.
This work was supported in part by the Swiss National Science Foundation.
This work was supported in part by the US Department of Energy under Award DE-FC02-04ER54698.